\documentclass[reprint,amsmath,amssymb,aps,prl]{revtex4-1}
\pdfoutput=1
\usepackage{times}
\usepackage{graphicx}

\begin{document}

\title{Fundamental Limits of Ultrathin Metasurfaces}

\author{Amir Arbabi}
\affiliation{T. J. Watson Laboratory of Applied Physics, California Institute of Technology, 1200 E California Blvd., Pasadena, CA 91125, USA}
\author{Andrei Faraon}
\email{Corresponding authors AA: amir@caltech.edu, AF: faraon@caltech.edu.}
\affiliation{T. J. Watson Laboratory of Applied Physics, California Institute of Technology, 1200 E California Blvd., Pasadena, CA 91125, USA}

\begin{abstract}
We present universal theoretical limits on the operation and performance of non-magnetic passive ultrathin metasurfaces. In particular, we prove that their local transmission, reflection, and polarization conversion coefficients are confined to limited regions of the complex plane. As a result, full control over the phase of the light transmitted through such metasurfaces cannot be achieved if the polarization of the light is not to be affected at the same time. We also establish fundamental limits on the maximum polarization conversion efficiency of these metasurfaces, and show that they cannot achieve more than 25\% polarization conversion efficiency in transmission.   
\end{abstract}
\maketitle

\section{Introduction}
Metamaterials are artificial materials with electromagnetic properties which are controllable by design. They are generally three dimensional periodic arrays of scatters with deep subwavelength periods, and their electromagnetic properties can be fully represented by their permittivity and permeability tensors. Metasurfaces are two dimensional counterparts of the metamaterials. They are composed of an array of  scatterers with sub-wavelength period which are located on a planar surface. Ultrathin metasurfaces are a class of metasurfaces composed of scatterers which are significantly thinner than the wavelength of the light~\cite{Kildishev2013a, Yu2014}. An ultrathin metasurface is generally created by subwavelength patterning of an ultrathin film. The film is usually deposited on a flat substrate, and is rationally patterned for modification of the phase~\cite{Huang2008,Yu2011,Aieta2012,Ni2013}, amplitude~\cite{Liu2014}, or polarization~\cite{Papakostas2003, Elliott2004, Zhao2011, Yu2012, Pors2011} of the transmitted or the reflected light. The main feature that distinguishes the ultrathin metasurfaces from conventional diffractive elements and other types of metastructures is their distinctive  principle of operation. Ultrathin metasurfaces cause a discontinuity in the phase of the light that is transmitted through or reflected from them. This phase discontinuity is the result of the interference between the incident wave and the scattered light by the resonant ultrathin scatterers~\cite{Yu2011,Aieta2012}. This is in contrast to the conventional diffractive elements and other types of metasructures which rely on gradual phase shifts accumulation during light propagation across them.  
Such metasurfaces have attracted a lot of attention recently and several types of flat diffractive elements such as lenses, axicons, and complex beam shapers~\cite{Aieta2012, Genevet2012, Karimi2014} have been realized using them. 

Materials with plasmonic resonances such as gold and silver are popular choices for the metasurface layer. This is because these materials have large refractive index values; therefore, even a thin layer of them can scatter the light significantly. One of the well-known drawbacks of using these materials is their substantial absorption loss which limits the efficiency of the diffractive elements. As a result, most of the work in this area have been limited to the near and mid-infrared, terahertz, and microwave wavelengths range where the absorption losses are smaller~\cite{Yu2014, Hu2013a, Liu2014, Pfeiffer2013}. Here, we present fundamental relations between the transmission, reflection, and polarization conversion coefficients of passive non-magnetic ultrathin metasurfaces. We show that theses fundamental relations limit the functionality and performance of these metasurfaces. As two example applications, we discuss the implications that these relations have on the phase front control and polarization modification using ultrathin metasurfaces. Some special cases of the limitations presented here have also been recently shown using network scattering matrix theory~\cite{Monticone2013}. 

\section{Fundamental Relations of Ultrathin Metasurfaces}
\begin{figure}[htp]
\centering
\includegraphics[width=\columnwidth]{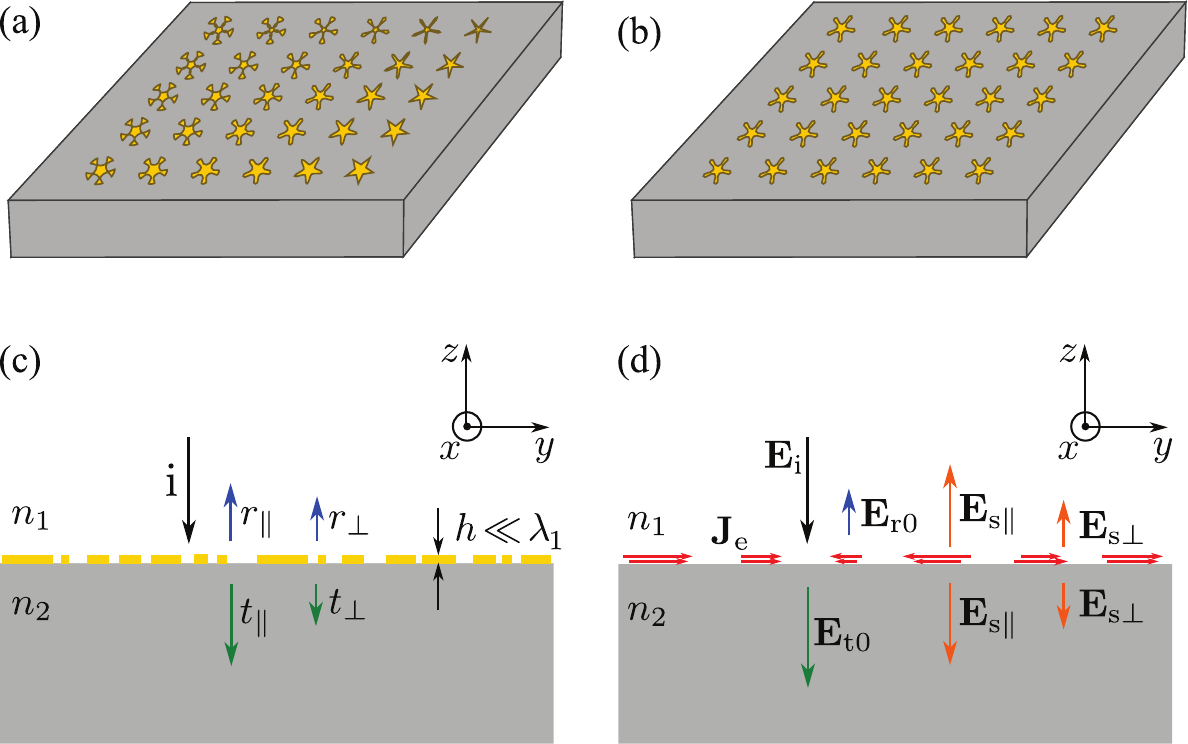}
\caption{(a) Schematic illustration of (a) a metasurface with gradually varying scatterers, and (b) a periodic metasurface. (c) Cross section view of the periodic metasurface shown in (b). A normally incident plane wave is impinging on the metasurface. The transmitted and reflected plane waves, and the polarization converted plane waves are also shown. (d) Illustration of the equivalent volume current density that has replaced the metasurface. $\mathbf{E}_\mathrm{i}$, $\mathbf{E}_\mathrm{r0}$, and $\mathbf{E}_\mathrm{t0}$ are the electric fields of the incident, reflected and transmitted plane waves for a bare interface, respectively. The electric field of the light scattered by the metasurface is labeled by $\mathbf{E}_\mathrm{s\parallel}$ for the part with the same polarization as $\mathbf{E}_\mathrm{r0}$ and $\mathbf{E}_\mathrm{t0}$, and by $\mathbf{E}_\mathrm{s\perp}$ for the part with polarization orthogonal to the polarizations of $\mathbf{E}_\mathrm{r0}$ and $\mathbf{E}_\mathrm{t0}$.}
\label{fig:metasurface_limit}
\end{figure}

We consider an ultrathin metasurface which is composed of an array of potentially different subwavelength passive and non-magnetic scatterers. A schematic illustration of such a metasurface is depicted in Fig.~\ref{fig:metasurface_limit}a. The scatterers are resting on a substrate with refractive index of $n_2$ and are surrounded by a cladding material with the refractive index of $n_1$. We assume that the scatterers have a thickness of $h$ which is much smaller than the wavelength of light in the cladding material (i.e. $h\!\ll\!\lambda_1\!=\!\lambda_0/n_1$). The metasurface locally modifies the amplitude, phase, or polarization of an incident light either in transmission or reflection. For gradually enough varying scatterers, the metasurface can be modelled as a surface with spatially dependent local reflection and transmission coefficients. 

For each of the scatterers comprising the metasurface, we can form a periodic metasurface by arranging that scatterer on a periodic lattice similar to the lattice of the original metasurface in the vicinity of that scatterer. An example of such a periodic metasurface is shown in Fig.~\ref{fig:metasurface_limit}b. The reflection and transmission of such a periodic metasurface approximates the local reflection and transmission of the original metasurface at the location of that scatterer. Thus, the properties that we establish for the reflection and transmission coefficients of the periodic are applicable to the local coefficients of the original metasurface.

We assume that a plane wave with a vacuum wavelength of $\lambda_0$ is normally incident on the periodic metasurface as shown in Fig.~\ref{fig:metasurface_limit}c. Since the period of the metasurface is smaller than the wavelength in both the surrounding materials, only the zeroth order transmission and reflection are propagating waves. In general, the transmitted and reflected light waves do not have the same polarization as the incident light. We decompose each of the transmitted and reflected plane waves into two plane waves with orthogonal polarizations. We represent the transmission and reflection coefficients for the parts of the light with the same polarizations as the plane waves transmitted through and reflected from a bare substrate-cladding interface (e.g. the interface without the metasurface) by $t_\parallel$ and $r_\parallel$, respectively. For a linearly polarized light, $t_\parallel$ and $r_\parallel$ represent the transmission and reflection coefficients of the parts of the light with the same polarization as the incident light. These coefficients are the total transmission and reflection coefficients for a metasurface that does not modify the polarization. Polarization modification happens if the light reflected from or transmitted through the metasurface acquire some polarization component orthogonal to the polarization of a plane wave reflected from or transmitted through the bare interface. We represent the transmission and reflection coefficients for the parts of the light whose polarization has been converted to the orthogonal polarization by $t_\perp$ and $r_\perp$, respectively. We refer to $t_\perp$ and $r_\perp$ as polarization conversion coefficients. For example, if the incident plane wave is right handed circularly polarized then $t_\parallel$ and $t_\perp$ are the transmission coefficients for the parts of the transmitted light which are right and left handed circularly polarized, respectively, while  $r_\parallel$ and $r_\perp$ are the reflection coefficients for the parts which are left and right handed circularly polarized, respectively.

Since the materials composing the periodic metasurface are non-magnetic, we can replace the entire metasurface by an equivalent volume electric current density 
\begin{equation}
\mathbf{J}_\mathrm{e}=-i\omega\epsilon_0(n_\mathrm{ms}^2-n_1^2)\mathbf{E},
\label{eq:volume_current}
\end{equation}
where $\omega$ is the angular frequency of the incident light, $\epsilon_0$ is the vacuum permittivity, $n_\mathrm{ms}$ is the complex refractive index of the metasurface materials, and $\mathbf{E}$ is the the total electric field~\cite{Balanis1989}. Note that $\mathbf{J}_\mathrm{e}$ is nonzero only at the location of the metasurface, and is confined to the thickness $h$ above the substrate as illustrated in Fig.~\ref{fig:metasurface_limit}d. According to the volume equivalence theorem, the light emitted by $\mathbf{J}_\mathrm{e}$ in the presence of the substrate-cladding interface is equal to the light scattered by the metasurface. The volume surface current density $\mathbf{J}_\mathrm{e}$ is also periodic with the same period as the periodic metasurface it has replaced; therefore, it emits into the substrate and top cladding materials only along the interface normal directions ($\pm$z directions in Figs.~\ref{fig:metasurface_limit}c and d).

To find the amplitudes of the plane waves emitted by $\mathbf{J}_\mathrm{e}$ into the cladding and substrate, we first find the amplitudes of the plane waves it emits when it is located in the material with the refractive index of $n_1$. For simplicity, we choose the coordinate system with $z=0$ plane located at the middle of the metasurface layer; therefore, $\mathbf{J}_\mathrm{e}$ is confined to $-h/2\!<\!z\!<\!h/2$ slab region. The electric fields of the plane waves emitted by $\mathbf{J}_\mathrm{e}$ along the $+z$ and $-z$ directions right above ($\mathbf{E}_{+z}$) and below ($\mathbf{E}_{-z}$) the slab region are given by~\cite{Collin1990}
\begin{align}\nonumber
\mathbf{E}_{\pm z}&=\frac{-Z_0}{2n_1}\mathrm{e}^{ik_1\frac{h}{2}}\iiint{\mathbf{J}_\mathrm{t}\mathrm{e}^{\mp ik_1z}\mathrm{d}x\mathrm{d}y\mathrm{d}z}\\ 
&=\frac{-Z_0}{2n_1}\mathrm{e}^{ik_1\frac{h}{2}}\int_{-\frac{h}{2}}^{\frac{h}{2}}{\mathbf{I}\mathrm{e}^{\mp ik_1z}\mathrm{d}z},
\label{eq:emitted_fields_homogenous}
\end{align}
where $k_1=n_12\pi/\lambda_0$, $Z_0$ is the impedance of free space, $\mathbf{J}_\mathrm{t}$ is the components of the $\mathbf{J}_\mathrm{e}$ parallel to the $xy$ plane, and $\mathbf{I}\triangleq\iint{\mathbf{J_\mathrm{t}}\mathrm{d}x\mathrm{d}y}$. Since $|k_1z|\!<\!\pi h/\lambda_1\!\ll\!1$, to the zeroth order in $h/\lambda_1$, the exponential terms in~(\ref{eq:emitted_fields_homogenous}) can be approximated by 1, and we obtain
\begin{equation}
\mathbf{E}_{\pm z}\approx\frac{-Z_0}{2n_1}\int_{-\frac{h}{2}}^{\frac{h}{2}}{\mathbf{I}\mathrm{d}z}.
\label{eq:approximation}
\end{equation}
According to~(\ref{eq:approximation}), if $\int_{-h/2}^{h/2}{\mathbf{I}\mathrm{d}z}$ is nonzero then the electric fields of the optical waves emitted by $\mathbf{J}_\mathrm{e}$ along the $\pm z$ directions are equal to each other. In the special case that $\int_{-h/2}^{h/2}{\mathbf{I}\mathrm{d}z}$ is zero, the first order term in the Taylor series expansion of $\mathbf{E}_{\pm z}$ would be the dominant term and $\mathbf{E}_{+z}$ and $\mathbf{E}_{-z}$ would not be equal to each other. For $\int_{-h/2}^{h/2}{\mathbf{I}\mathrm{d}z}$ to become zero, the phase of $\mathbf{I}$ and thus $\mathbf{J}_\mathrm{e}$ should vary at least by $\pi$ as a function of $z$ within the thickness of the metasurface. As we mentioned earlier, the light does not accumulate significant phase as it propagates through an ultrathin metasurfaces which operates based on creating a phase discontinuity. Therefore, the phase of the electric field $\mathbf{E}$ inside the metasurface layer and, according to~(\ref{eq:volume_current}), the phase of $\mathbf{J}_\mathrm{e}$ does not vary significantly along the propagation direction inside an ultrathin metasurface. As a result, $\int_{-h/2}^{h/2}{\mathbf{I}\mathrm{d}z}$ would be nonzero and the amplitudes of the plane waves emitted by $\mathbf{J}_\mathrm{e}$ toward the $\pm z$ directions would be equal to each other. 

Some of the light emitted by $\mathbf{J}_\mathrm{e}$ toward the $-z$ direction is reflected back at the substrate-cladding interface. We obtain the electric field amplitudes of the plane waves emitted by $\mathbf{J}_\mathrm{e}$ inside the substrate ($\mathbf{E}_\mathrm{s-}$) and cladding ($\mathbf{E}_\mathrm{s+}$) as 
\begin{subequations}
\begin{align}
\mathbf{E}_\mathrm{s-}&=t_0\mathbf{E}_{-z},\\
\mathbf{E}_\mathrm{s+}&=\mathbf{E}_{+z}+\mathrm{e}^{ik_1h}r_0\mathbf{E}_{-z}\approx(1+r_0)\mathbf{E}_{-z}=\mathbf{E}_\mathrm{s-},
\end{align}
\label{eq:scattered_fields}
\end{subequations}
where $r_0=(n_1-n_2)/(n_1+n_2)$ and $t_0=2n_1/(n_1+n_2)$  are respectively the reflection and transmission coefficients of the bare substrate-cladding interface for normally incident plane waves. In~(\ref{eq:scattered_fields})b we have used   $\exp(ik_1h)\approx1$  and the relation $t_0\!=\!1+r_0$. Since the the light emitted by $\mathbf{J}_\mathrm{e}$  into the substrate and cladding layers are equal to each other we omit the + and - subscripts and represent both by $\mathbf{E}_\mathrm{s}$. According to the superposition principle~\cite{Balanis1989}, the total transmitted and reflected optical waves are the summation of the waves transmitted  through  and reflected from the bare substrate-cladding interface, plus the light emitted by $\mathbf{J}_\mathrm{e}$. As shown in Fig.~\ref{fig:metasurface_limit}d, $\mathbf{E}_\mathrm{s}$ is decomposed into two plane waves with orthogonal polarizations. The parts of $\mathbf{E}_\mathrm{s}$ which have the same polarization as $\mathbf{E}_\mathrm{t_0}$ and $\mathbf{E}_\mathrm{r_0}$  are represented by $\mathbf{E}_\mathrm{s\parallel}$, and the parts which have polarizations orthogonal to them are shown by $\mathbf{E}_\mathrm{s\perp}$. 

Since the periodic metasurface is passive the sum of the transmitted and reflected powers is equal to, or smaller than (for lossy metasurfaces) the incident power, that is
\begin{equation}
|t_\parallel|^2+|r_\parallel|^2+|t_\perp|^2+|r_\perp|^2\leq1.
\label{eq:power_conservation}
\end{equation}
For the bare interface we have
\begin{equation}
\mathbf{E}_\mathrm{t_0}=\mathbf{E}_\mathrm{i}+\mathbf{E}_\mathrm{r_0}.
\label{eq:bare_interface_relation}
\end{equation}
We can express the transmission and reflection coefficients of the periodic metasurface in terms of the electric field amplitudes as
\begin{subequations}
\begin{align}
&t_\parallel=\sqrt{\frac{n_2}{n_1}}\mathrm{\frac{E_{t_0}+E_{s\parallel}}{E_i},}\\
&r_\parallel=\mathrm{\frac{E_{r_0}+E_{s\parallel}}{E_i},}\\
&t_\perp=\sqrt{\frac{n_2}{n_1}}\mathrm{\frac{E_{s\perp}}{E_i},}\\
&r_\perp=\mathrm{\frac{E_{s\perp}}{E_i}.}
\end{align}\label{eq:tras_ref_fields}
\end{subequations}

Using~(\ref{eq:power_conservation}),~(\ref{eq:bare_interface_relation}), and~(\ref{eq:tras_ref_fields}) we find the fundamental relations of metasurfaces as
\begin{subequations}
\begin{align}\label{eq:fundamental_limit_a}
&|t_\parallel|^2+|r_\parallel|^2+|t_\perp|^2+|r_\perp|^2\leq1,\\\label{eq:fundamental_limit_b}
&r_\parallel=\sqrt{\frac{n_1}{n_2}}t_\parallel-1,\\\label{eq:fundamental_limit_c}
&r_\perp=\sqrt{\frac{n_1}{n_2}}t_\perp.
\end{align}\label{eq:fundamental_limit}
\end{subequations}
By using a similar procedure we can show that when the incident light is incident at an angle $\theta_i$ with respect to the interface normal direction, relation ~(\ref{eq:fundamental_limit_a}) is still valid and metasurface relations~(\ref{eq:fundamental_limit_b}) and~(\ref{eq:fundamental_limit_c}) are modified as
\begin{subequations}
\begin{align}
&r_\parallel=\sqrt{\frac{n_1\cos(\theta_i)}{n_2\cos(\theta_r)}}t_\parallel-1,\\
&r_\perp=\sqrt{\frac{n_1\cos(\theta_r)}{n_2\cos(\theta_i)}}t_\perp,
\end{align}\label{eq:fundamental_limit_TE}
\end{subequations}
for a transverse electric (TE) polarized incident plane wave, and as 
\begin{subequations}
\begin{align}
&r_\parallel=\sqrt{\frac{n_1\cos(\theta_r)}{n_2\cos(\theta_i)}}t_\parallel-1,\\
&r_\perp=\sqrt{\frac{n_1\cos(\theta_i)}{n_2\cos(\theta_r)}}t_\perp.
\end{align}\label{eq:fundamental_limit_TM}
\end{subequations}
for a transverse electric (TM) polarized incident light. Here, $\theta_r$ is the angle of refraction and $\cos(\theta_r)=\sqrt{1-(n_1/n_2)^2\sin^2(\theta_i)}$.

The transmission and reflection coefficients of any periodic ultrathin metasurface satisfy~(\ref{eq:fundamental_limit}). As we mentioned earlier, the local transmission and reflection coefficients of a metasurface with gradually enough varying parameters is similar to those of a periodic metasurface without the gradual variations of the metasurface. Therefore, the relations~(\ref{eq:fundamental_limit}) are valid for the local transmission, reflection and polarization conversion coefficients of a metasurface.  It should be noted that the limitations expressed by~(\ref{eq:fundamental_limit}) are merely the results of deep subwavelength thickness of the metasurface, and are valid regardless of the possible absorption loss caused by metasurfaces. Material absorption loss will tighten the limit in ~(\ref{eq:fundamental_limit_a})  even further. More specifically, for a lossy metasurface the left hand side of~(\ref{eq:fundamental_limit_a}) is equal to $1-L$, where $L$ is the fraction of the light absorbed by the metasurface.
 
 In the followings, as two example cases, we discuss the implications that the relations~(\ref{eq:fundamental_limit}) have on the performance of reflective and transmissive metasurfaces designed for shaping the phase front, and modifying the polarization of a normally incident light. 

\subsection{Phase Front Modification with Ultrathin Metasurfaces}
Consider an ultrathin metasurface which is designed to shape the phase front of an incident beam to a desired form without disturbing its polarization. For such a metasurface, by using~(\ref{eq:fundamental_limit})a and~(\ref{eq:fundamental_limit})b, we find
\begin{equation}
|t_\parallel|^2+|\sqrt{\frac{n_1}{n_2}}t_\parallel-1|^2\leq1-|t_\perp|^2-|r_\perp|^2\leq1.
\label{eq:co_transmission}
\end{equation}
We can also express~(\ref{eq:co_transmission}) in terms of reflection coefficient as
\begin{equation}
\frac{n_2}{n_1}|1+r_\parallel|^2+|r_\parallel|^2\leq1.
\label{eq:co_reflecton}
\end{equation}
\begin{figure}[htp]
\centering
\includegraphics[width=0.65\columnwidth]{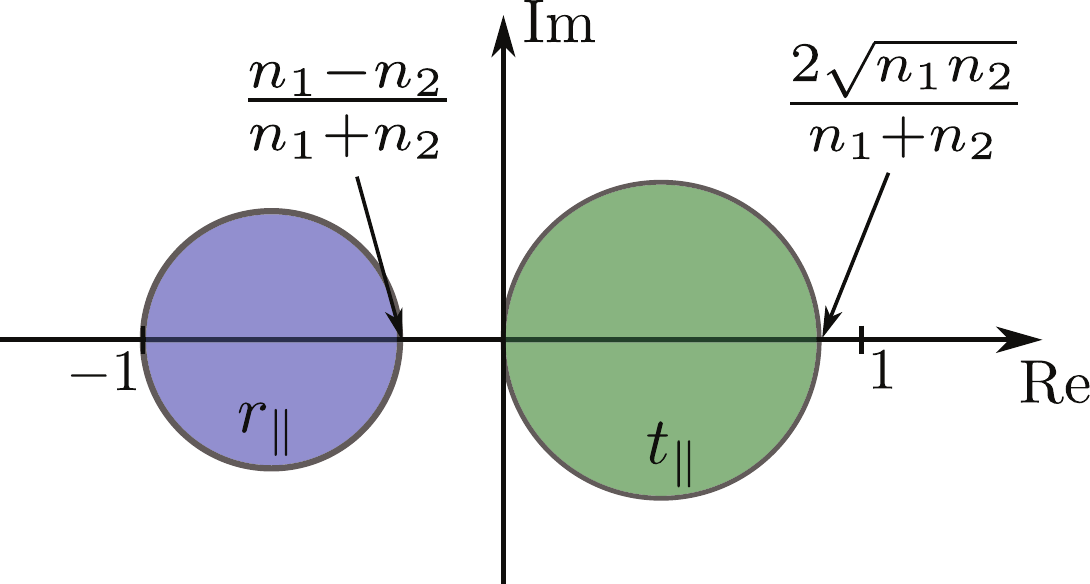}
\caption{Accessible regions of the complex plane for the transmission and reflection coefficients of a metasurface for $n_2>n_1$.}
\label{fig:metasurface_limit_phase_results}
\end{figure}

The inequalities~(\ref{eq:co_transmission}) and~(\ref{eq:co_reflecton}) limit the transmission and reflection coefficients to two circles in the complex plane as shown in Fig.~\ref{fig:metasurface_limit_phase_results}. As we can see from Fig.~\ref{fig:metasurface_limit_phase_results}, the phase of the transmission coefficient is limited to the $(-\pi/2,\pi/2)$ interval. Therefore, a transmissive ultrathin metasurface cannot provide full control over the phase of the transmitted light which has the same polarization as the incident light. When $n_1\leq n_2$ (i.e. the light is incident from the material with the lower refractive index), the attainable reflection coefficients also only cover a phase range smaller than $\pi$ and therefore an arbitrary reflective phase mask cannot be implemented using a metasurface. When the light is incident from the material with larger refractive index (i.e. $n_1\!>\!n_2$), the phase of the reflection coefficient might cover the full $2\pi$ range, but if $n_1$ and $n_2$ are of the same order of magnitude then the reflection efficiency cannot be large for all phases. For example, if the ultrathin metasurface is fabricated on a silicon substrate with refractive index of $n_1=3.48$ and cladded with air, then the reflection efficiency of an infrared light which is incident from the silicon side and is reflected with zero phase is smaller than $(n_1-n_2)^2/(n_1+n_2)^2=31\%$.

\subsection{Polarization Manipulation using Ultrathin Metasurfaces}
Ultrathin metasurfaces can also be designed to modify the polarization of an incident light. Polarization modification is achieved by converting the polarization of all or part of the incident light to a polarization orthogonal to that of the incident light. For such metasurfaces, by using~(\ref{eq:fundamental_limit}) we obtain
\begin{equation}
|t_\perp|^2(1+\frac{n_1}{n_2})\leq1-|t_\parallel|^2-|\sqrt{\frac{n_1}{n_2}}t_\parallel-1|^2.
\label{eq:cross_trans_limit_1}
\end{equation}
The right hand side of~(\ref{eq:cross_trans_limit_1}) can be simplified further as 
\begin{align}\nonumber
&1-|t_\parallel|^2-|\sqrt{\frac{n_1}{n_2}}t_\parallel-1|^2=2\sqrt{\frac{n_1}{n_2}}\mathrm{Re}\{t_\parallel\}-(1+\frac{n_1}{n_2})|t_\parallel|^2\\
&\leq2\sqrt{\frac{n_1}{n_2}}|t_\parallel|-(1+\frac{n_1}{n_2})|t_\parallel|^2\leq\frac{n_1}{(n_1+n_2)}.
\label{eq:cross_trans_limit_2}
\end{align}
The left hand side of the last inequality in~(\ref{eq:cross_trans_limit_2}) is a quadratic function of $|t_\parallel|$, and the right hand side of the inequality represents the maximum of this quadratic function. By combining~(\ref{eq:cross_trans_limit_1}) and~(\ref{eq:cross_trans_limit_2}) we find the limit on the polarization conversion efficiency of the transmitted light as
\begin{equation}
|t_\perp|^2\leq\frac{n_1n_2}{(n_1+n_2)^2},
\label{eq:cross_transmission}
\end{equation}
and by using~(\ref{eq:fundamental_limit})c we obtain the limit on the polarization conversion efficiency in reflection as 
\begin{equation}
|r_\perp|^2\leq\frac{n_1^2}{(n_1+n_2)^2}.
\label{eq:cross_reflection}
\end{equation}

\begin{figure}[htp]
\centering
\includegraphics[width=0.9\columnwidth]{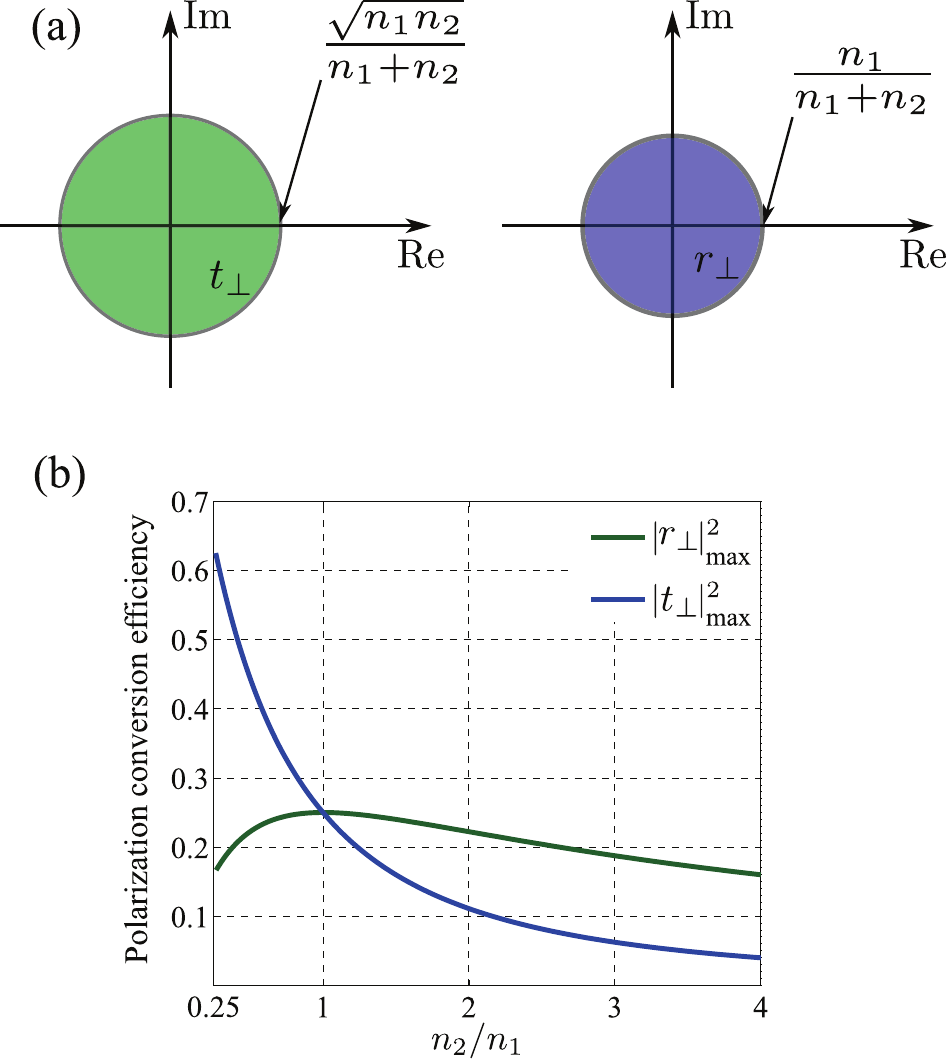}
\caption{(a) Regions of the complex plane admissible for the polarization conversion coefficients  $t_\perp$ and $r_\perp$. (b) Maximum polarization conversion efficiency for a transmissive and reflective ultrathin metasurface as a function of the ratio of the refractive index of the substrate to that of the cladding.}
\label{fig:metasurface_limit_polar_results}
\end{figure}

The inequalities~(\ref{eq:cross_transmission}) and~(\ref{eq:cross_reflection}) limit the performance of a metasurface designed to manipulate the light's polarization. For example, the maximum efficiency of a transmissive metasurface half-wave plate which rotates the polarization of a linearly polarized beam by 90 degrees, and is fabricated on a fused silica substrate with the refractive index of $n_2=1.44$ is about 24\%. As another example, a transmissive metasurface quarter-wave plate fabricated on the fused silica substrate cannot be more than 48\% efficient in converting the polarization from linear to circular. Similar limits can be established for efficiencies of any metasurface that modifies the polarization of light.  Figure~\ref{fig:metasurface_limit_polar_results}a shows the values in the complex plane that $t_\perp$ and $r_\perp$ can achieve. As this figure shows, the phase of $t_\perp$ and $r_\perp$  may take any values in the $(0,2\pi)$ interval, but the transmission and reflection efficiencies are limited to the values enforced by~(\ref{eq:cross_transmission}) and~(\ref{eq:cross_reflection}), respectively. Therefore, a limited efficiency transmissive phase mask which imposes an arbitrary phase profile might be implemented using a metasurface if we allow the polarization of the phase shifted transmitted light to be orthogonal to that of the incident light. Such metasurfaces which implement phase masks corresponding to a lens and an axicon have been previously reported~\cite{Aieta2012}.

The maximum polarization conversion efficiencies for a metasurface as a function of the ratio of the substrate to cladding refractive indices has been plotted in Fig.~\ref{fig:metasurface_limit_polar_results}b. As this figure shows, a single transmissive metasurface cannot change the polarization with an efficiency more than 25\%. However, the efficiency of a single layer reflective metasurface can be higher if the light is impinging on the metasurface from the higher index material.  

\section{Beyond the Ultrathin Metasurface Limits}
The limitations we presented here are the results of the non-directionality of the scattering by the ultrathin metasurfaces. In other words, the scattered electric field amplitude of an ultrathin metasurface is equal in both the substrate and the cladding layers. These limitations may be overcome by using a multilayer metasurface as it has been shown in the microwave regime~\cite{McGrath1986, Pozar1996a, Encinar2003, Ryan2010, Pfeiffer2013}, or by adding an interface between two materials in a different layer as the ultrathin metasurface~\cite{Pozar1997, Pozar1999, Pors2013}. 

Another approach is to use thicker scatterers. Scatterers with the thickness on the order of a wavelength may scatter the light with unequal electric field amplitudes into the substrate and cladding layers; therefore, their operations are not limited by the fundamental ultrathin metasurface relations. If we use lossy materials such as gold or silver then thick scatterers absorb significantly, and the absorption limits the efficiency of such metastructures~\cite{Verslegers2009}. In this case, the solution is to replace the metal with a low loss dielectric. Indeed, achieving performances beyond the ultrathin metasurface limits have been recently achieved using 1D high contrast gratings~\cite{Fattal2010,Lu2010} and high contrast transmitarrays~\cite{Vo2014,Arbabi2014,Arbabi2014c,Arbabi2014a,Arbabi2014d}. High contrast tranmsitarrays are composed of a two dimensional dissimilar dielectric scatterers with the thickness on the order of a wavelength which are arranged on a periodic lattice. Highly efficient full phase control in transmission~\cite{Vo2014,Arbabi2014, Arbabi2014c,Arbabi2014d} and reflection~\cite{Arbabi2014a} without altering the polarization (which are fundamentally unachievable using metasurfaces) have been demonstrated.

\section*{Conclusion}
We showed that the local transmission, reflection and polarization conversion coefficients of a non-magnetic passive ultrathin metasurface satisfy a set of fundamental relations. These fundamental relations enforce some theoretical limitations on the complex values of these coefficients. Particularly, we demonstrated that an ultrathin metasurface cannot provide full phase control for implementation of an arbitrary phase mask unless it changes the  polarization of the light as well, and even in that case the efficiency of the phase mask is limited by the transmission polarization conversion efficiency and is smaller than 25\%. We can use two approaches to overcome the limitations of ultrathin single layer metasurfaces; cascading ultrathin metasurfaces with other ultrathin metasurfaces or planar interfaces, or using high contrast transmitarrays. The latter not only is not limited by the ultrathin metasurface fundamental relations but also does not suffer from the material absorption loss.

\section*{Acknowledgement}

This work was supported by Caltech/JPL president and director fund (PDF). The authors would like to thank Dr. David Fattal, Dr. Nader Engheta, and Dr. Shanhui Fan for useful discussions.


\begin{thebibliography}{33}%
\makeatletter
\providecommand \@ifxundefined [1]{%
 \@ifx{#1\undefined}
}%
\providecommand \@ifnum [1]{%
 \ifnum #1\expandafter \@firstoftwo
 \else \expandafter \@secondoftwo
 \fi
}%
\providecommand \@ifx [1]{%
 \ifx #1\expandafter \@firstoftwo
 \else \expandafter \@secondoftwo
 \fi
}%
\providecommand \natexlab [1]{#1}%
\providecommand \enquote  [1]{``#1''}%
\providecommand \bibnamefont  [1]{#1}%
\providecommand \bibfnamefont [1]{#1}%
\providecommand \citenamefont [1]{#1}%
\providecommand \href@noop [0]{\@secondoftwo}%
\providecommand \href [0]{\begingroup \@sanitize@url \@href}%
\providecommand \@href[1]{\@@startlink{#1}\@@href}%
\providecommand \@@href[1]{\endgroup#1\@@endlink}%
\providecommand \@sanitize@url [0]{\catcode `\\12\catcode `\$12\catcode
  `\&12\catcode `\#12\catcode `\^12\catcode `\_12\catcode `\%12\relax}%
\providecommand \@@startlink[1]{}%
\providecommand \@@endlink[0]{}%
\providecommand \url  [0]{\begingroup\@sanitize@url \@url }%
\providecommand \@url [1]{\endgroup\@href {#1}{\urlprefix }}%
\providecommand \urlprefix  [0]{URL }%
\providecommand \Eprint [0]{\href }%
\providecommand \doibase [0]{http://dx.doi.org/}%
\providecommand \selectlanguage [0]{\@gobble}%
\providecommand \bibinfo  [0]{\@secondoftwo}%
\providecommand \bibfield  [0]{\@secondoftwo}%
\providecommand \translation [1]{[#1]}%
\providecommand \BibitemOpen [0]{}%
\providecommand \bibitemStop [0]{}%
\providecommand \bibitemNoStop [0]{.\EOS\space}%
\providecommand \EOS [0]{\spacefactor3000\relax}%
\providecommand \BibitemShut  [1]{\csname bibitem#1\endcsname}%
\let\auto@bib@innerbib\@empty
\bibitem [{\citenamefont {Kildishev}\ \emph {et~al.}(2013)\citenamefont
  {Kildishev}, \citenamefont {Boltasseva},\ and\ \citenamefont
  {Shalaev}}]{Kildishev2013a}%
  \BibitemOpen
  \bibfield  {author} {\bibinfo {author} {\bibfnamefont {A.~V.}\ \bibnamefont
  {Kildishev}}, \bibinfo {author} {\bibfnamefont {A.}~\bibnamefont
  {Boltasseva}}, \ and\ \bibinfo {author} {\bibfnamefont {V.~M.}\ \bibnamefont
  {Shalaev}},\ }\href {\doibase 10.1126/science.1232009} {\bibfield  {journal}
  {\bibinfo  {journal} {Science (New York, N.Y.)}\ }\textbf {\bibinfo {volume}
  {339}},\ \bibinfo {pages} {1232009} (\bibinfo {year} {2013})}\BibitemShut
  {NoStop}%
\bibitem [{\citenamefont {Yu}\ and\ \citenamefont {Capasso}(2014)}]{Yu2014}%
  \BibitemOpen
  \bibfield  {author} {\bibinfo {author} {\bibfnamefont {N.}~\bibnamefont
  {Yu}}\ and\ \bibinfo {author} {\bibfnamefont {F.}~\bibnamefont {Capasso}},\
  }\href {\doibase 10.1038/NMAT3839} {\bibfield  {journal} {\bibinfo  {journal}
  {Nature materials}\ }\textbf {\bibinfo {volume} {13}},\ \bibinfo {pages}
  {139} (\bibinfo {year} {2014})}\BibitemShut {NoStop}%
\bibitem [{\citenamefont {Huang}\ \emph {et~al.}(2008)\citenamefont {Huang},
  \citenamefont {Kao}, \citenamefont {Fedotov}, \citenamefont {Chen},\ and\
  \citenamefont {Zheludev}}]{Huang2008}%
  \BibitemOpen
  \bibfield  {author} {\bibinfo {author} {\bibfnamefont {F.~M.}\ \bibnamefont
  {Huang}}, \bibinfo {author} {\bibfnamefont {T.~S.}\ \bibnamefont {Kao}},
  \bibinfo {author} {\bibfnamefont {V.~a.}\ \bibnamefont {Fedotov}}, \bibinfo
  {author} {\bibfnamefont {Y.}~\bibnamefont {Chen}}, \ and\ \bibinfo {author}
  {\bibfnamefont {N.~I.}\ \bibnamefont {Zheludev}},\ }\href {\doibase
  10.1021/nl801476v} {\bibfield  {journal} {\bibinfo  {journal} {Nano letters}\
  }\textbf {\bibinfo {volume} {8}},\ \bibinfo {pages} {2469} (\bibinfo {year}
  {2008})}\BibitemShut {NoStop}%
\bibitem [{\citenamefont {Yu}\ \emph {et~al.}(2011)\citenamefont {Yu},
  \citenamefont {Genevet}, \citenamefont {Kats}, \citenamefont {Aieta},
  \citenamefont {Tetienne}, \citenamefont {Capasso},\ and\ \citenamefont
  {Gaburro}}]{Yu2011}%
  \BibitemOpen
  \bibfield  {author} {\bibinfo {author} {\bibfnamefont {N.}~\bibnamefont
  {Yu}}, \bibinfo {author} {\bibfnamefont {P.}~\bibnamefont {Genevet}},
  \bibinfo {author} {\bibfnamefont {M.~a.}\ \bibnamefont {Kats}}, \bibinfo
  {author} {\bibfnamefont {F.}~\bibnamefont {Aieta}}, \bibinfo {author}
  {\bibfnamefont {J.-P.}\ \bibnamefont {Tetienne}}, \bibinfo {author}
  {\bibfnamefont {F.}~\bibnamefont {Capasso}}, \ and\ \bibinfo {author}
  {\bibfnamefont {Z.}~\bibnamefont {Gaburro}},\ }\href {\doibase
  10.1126/science.1210713} {\bibfield  {journal} {\bibinfo  {journal} {Science
  (New York, N.Y.)}\ }\textbf {\bibinfo {volume} {334}},\ \bibinfo {pages}
  {333} (\bibinfo {year} {2011})}\BibitemShut {NoStop}%
\bibitem [{\citenamefont {Aieta}\ \emph {et~al.}(2012)\citenamefont {Aieta},
  \citenamefont {Genevet}, \citenamefont {Kats}, \citenamefont {Yu},
  \citenamefont {Blanchard}, \citenamefont {Gaburro},\ and\ \citenamefont
  {Capasso}}]{Aieta2012}%
  \BibitemOpen
  \bibfield  {author} {\bibinfo {author} {\bibfnamefont {F.}~\bibnamefont
  {Aieta}}, \bibinfo {author} {\bibfnamefont {P.}~\bibnamefont {Genevet}},
  \bibinfo {author} {\bibfnamefont {M.~A.}\ \bibnamefont {Kats}}, \bibinfo
  {author} {\bibfnamefont {N.}~\bibnamefont {Yu}}, \bibinfo {author}
  {\bibfnamefont {R.}~\bibnamefont {Blanchard}}, \bibinfo {author}
  {\bibfnamefont {Z.}~\bibnamefont {Gaburro}}, \ and\ \bibinfo {author}
  {\bibfnamefont {F.}~\bibnamefont {Capasso}},\ }\href {\doibase
  10.1021/nl302516v} {\bibfield  {journal} {\bibinfo  {journal} {Nano letters}\
  }\textbf {\bibinfo {volume} {12}},\ \bibinfo {pages} {4932} (\bibinfo {year}
  {2012})}\BibitemShut {NoStop}%
\bibitem [{\citenamefont {Ni}\ \emph {et~al.}(2013)\citenamefont {Ni},
  \citenamefont {Ishii}, \citenamefont {Kildishev},\ and\ \citenamefont
  {Shalaev}}]{Ni2013}%
  \BibitemOpen
  \bibfield  {author} {\bibinfo {author} {\bibfnamefont {X.}~\bibnamefont
  {Ni}}, \bibinfo {author} {\bibfnamefont {S.}~\bibnamefont {Ishii}}, \bibinfo
  {author} {\bibfnamefont {A.~V.}\ \bibnamefont {Kildishev}}, \ and\ \bibinfo
  {author} {\bibfnamefont {V.~M.}\ \bibnamefont {Shalaev}},\ }\href {\doibase
  10.1038/lsa.2013.28} {\bibfield  {journal} {\bibinfo  {journal} {Light:
  Science \& Applications}\ }\textbf {\bibinfo {volume} {2}},\ \bibinfo {pages}
  {e72} (\bibinfo {year} {2013})}\BibitemShut {NoStop}%
\bibitem [{\citenamefont {Liu}\ \emph {et~al.}(2014)\citenamefont {Liu},
  \citenamefont {Zhang}, \citenamefont {Kenney}, \citenamefont {Su},
  \citenamefont {Xu}, \citenamefont {Ouyang}, \citenamefont {Shi},
  \citenamefont {Han}, \citenamefont {Zhang},\ and\ \citenamefont
  {Zhang}}]{Liu2014}%
  \BibitemOpen
  \bibfield  {author} {\bibinfo {author} {\bibfnamefont {L.}~\bibnamefont
  {Liu}}, \bibinfo {author} {\bibfnamefont {X.}~\bibnamefont {Zhang}}, \bibinfo
  {author} {\bibfnamefont {M.}~\bibnamefont {Kenney}}, \bibinfo {author}
  {\bibfnamefont {X.}~\bibnamefont {Su}}, \bibinfo {author} {\bibfnamefont
  {N.}~\bibnamefont {Xu}}, \bibinfo {author} {\bibfnamefont {C.}~\bibnamefont
  {Ouyang}}, \bibinfo {author} {\bibfnamefont {Y.}~\bibnamefont {Shi}},
  \bibinfo {author} {\bibfnamefont {J.}~\bibnamefont {Han}}, \bibinfo {author}
  {\bibfnamefont {W.}~\bibnamefont {Zhang}}, \ and\ \bibinfo {author}
  {\bibfnamefont {S.}~\bibnamefont {Zhang}},\ }\href {\doibase
  10.1002/adma.201401484} {\bibfield  {journal} {\bibinfo  {journal} {Advanced
  Materials}\ ,\ \bibinfo {pages} {n/a}} (\bibinfo {year} {2014})}\BibitemShut
  {NoStop}%
\bibitem [{\citenamefont {Papakostas}\ \emph {et~al.}(2003)\citenamefont
  {Papakostas}, \citenamefont {Potts}, \citenamefont {Bagnall}, \citenamefont
  {Prosvirnin}, \citenamefont {Coles},\ and\ \citenamefont
  {Zheludev}}]{Papakostas2003}%
  \BibitemOpen
  \bibfield  {author} {\bibinfo {author} {\bibfnamefont {A.}~\bibnamefont
  {Papakostas}}, \bibinfo {author} {\bibfnamefont {A.}~\bibnamefont {Potts}},
  \bibinfo {author} {\bibfnamefont {D.}~\bibnamefont {Bagnall}}, \bibinfo
  {author} {\bibfnamefont {S.}~\bibnamefont {Prosvirnin}}, \bibinfo {author}
  {\bibfnamefont {H.}~\bibnamefont {Coles}}, \ and\ \bibinfo {author}
  {\bibfnamefont {N.}~\bibnamefont {Zheludev}},\ }\href {\doibase
  10.1103/PhysRevLett.90.107404} {\bibfield  {journal} {\bibinfo  {journal}
  {Physical Review Letters}\ }\textbf {\bibinfo {volume} {90}},\ \bibinfo
  {pages} {107404} (\bibinfo {year} {2003})}\BibitemShut {NoStop}%
\bibitem [{\citenamefont {Elliott}\ \emph {et~al.}(2004)\citenamefont
  {Elliott}, \citenamefont {Smolyaninov}, \citenamefont {Zheludev},\ and\
  \citenamefont {Zayats}}]{Elliott2004}%
  \BibitemOpen
  \bibfield  {author} {\bibinfo {author} {\bibfnamefont {J.}~\bibnamefont
  {Elliott}}, \bibinfo {author} {\bibfnamefont {I.}~\bibnamefont
  {Smolyaninov}}, \bibinfo {author} {\bibfnamefont {N.}~\bibnamefont
  {Zheludev}}, \ and\ \bibinfo {author} {\bibfnamefont {A.}~\bibnamefont
  {Zayats}},\ }\href {\doibase 10.1103/PhysRevB.70.233403} {\bibfield
  {journal} {\bibinfo  {journal} {Physical Review B}\ }\textbf {\bibinfo
  {volume} {70}},\ \bibinfo {pages} {233403} (\bibinfo {year}
  {2004})}\BibitemShut {NoStop}%
\bibitem [{\citenamefont {Zhao}\ and\ \citenamefont
  {Al\`{u}}(2011)}]{Zhao2011}%
  \BibitemOpen
  \bibfield  {author} {\bibinfo {author} {\bibfnamefont {Y.}~\bibnamefont
  {Zhao}}\ and\ \bibinfo {author} {\bibfnamefont {A.}~\bibnamefont {Al\`{u}}},\
  }\href {\doibase 10.1103/PhysRevB.84.205428} {\bibfield  {journal} {\bibinfo
  {journal} {Physical Review B}\ }\textbf {\bibinfo {volume} {84}},\ \bibinfo
  {pages} {205428} (\bibinfo {year} {2011})}\BibitemShut {NoStop}%
\bibitem [{\citenamefont {Yu}\ \emph {et~al.}(2012)\citenamefont {Yu},
  \citenamefont {Aieta}, \citenamefont {Genevet}, \citenamefont {Kats},
  \citenamefont {Gaburro},\ and\ \citenamefont {Capasso}}]{Yu2012}%
  \BibitemOpen
  \bibfield  {author} {\bibinfo {author} {\bibfnamefont {N.}~\bibnamefont
  {Yu}}, \bibinfo {author} {\bibfnamefont {F.}~\bibnamefont {Aieta}}, \bibinfo
  {author} {\bibfnamefont {P.}~\bibnamefont {Genevet}}, \bibinfo {author}
  {\bibfnamefont {M.~a.}\ \bibnamefont {Kats}}, \bibinfo {author}
  {\bibfnamefont {Z.}~\bibnamefont {Gaburro}}, \ and\ \bibinfo {author}
  {\bibfnamefont {F.}~\bibnamefont {Capasso}},\ }\href {\doibase
  10.1021/nl303445u} {\bibfield  {journal} {\bibinfo  {journal} {Nano letters}\
  }\textbf {\bibinfo {volume} {12}},\ \bibinfo {pages} {6328} (\bibinfo {year}
  {2012})}\BibitemShut {NoStop}%
\bibitem [{\citenamefont {Pors}\ \emph {et~al.}(2011)\citenamefont {Pors},
  \citenamefont {Nielsen}, \citenamefont {{Della Valle}}, \citenamefont
  {Willatzen}, \citenamefont {Albrektsen},\ and\ \citenamefont
  {Bozhevolnyi}}]{Pors2011}%
  \BibitemOpen
  \bibfield  {author} {\bibinfo {author} {\bibfnamefont {A.}~\bibnamefont
  {Pors}}, \bibinfo {author} {\bibfnamefont {M.~G.}\ \bibnamefont {Nielsen}},
  \bibinfo {author} {\bibfnamefont {G.}~\bibnamefont {{Della Valle}}}, \bibinfo
  {author} {\bibfnamefont {M.}~\bibnamefont {Willatzen}}, \bibinfo {author}
  {\bibfnamefont {O.}~\bibnamefont {Albrektsen}}, \ and\ \bibinfo {author}
  {\bibfnamefont {S.~I.}\ \bibnamefont {Bozhevolnyi}},\ }\href {\doibase
  10.1364/OL.36.001626} {\bibfield  {journal} {\bibinfo  {journal} {Optics
  letters}\ }\textbf {\bibinfo {volume} {36}},\ \bibinfo {pages} {1626}
  (\bibinfo {year} {2011})}\BibitemShut {NoStop}%
\bibitem [{\citenamefont {Genevet}\ \emph {et~al.}(2012)\citenamefont
  {Genevet}, \citenamefont {Yu}, \citenamefont {Aieta}, \citenamefont {Lin},
  \citenamefont {Kats}, \citenamefont {Blanchard}, \citenamefont {Scully},
  \citenamefont {Gaburro},\ and\ \citenamefont {Capasso}}]{Genevet2012}%
  \BibitemOpen
  \bibfield  {author} {\bibinfo {author} {\bibfnamefont {P.}~\bibnamefont
  {Genevet}}, \bibinfo {author} {\bibfnamefont {N.}~\bibnamefont {Yu}},
  \bibinfo {author} {\bibfnamefont {F.}~\bibnamefont {Aieta}}, \bibinfo
  {author} {\bibfnamefont {J.}~\bibnamefont {Lin}}, \bibinfo {author}
  {\bibfnamefont {M.~A.}\ \bibnamefont {Kats}}, \bibinfo {author}
  {\bibfnamefont {R.}~\bibnamefont {Blanchard}}, \bibinfo {author}
  {\bibfnamefont {M.~O.}\ \bibnamefont {Scully}}, \bibinfo {author}
  {\bibfnamefont {Z.}~\bibnamefont {Gaburro}}, \ and\ \bibinfo {author}
  {\bibfnamefont {F.}~\bibnamefont {Capasso}},\ }\href {\doibase
  10.1063/1.3673334} {\bibfield  {journal} {\bibinfo  {journal} {Applied
  Physics Letters}\ }\textbf {\bibinfo {volume} {100}},\ \bibinfo {pages}
  {013101} (\bibinfo {year} {2012})}\BibitemShut {NoStop}%
\bibitem [{\citenamefont {Karimi}\ \emph {et~al.}(2014)\citenamefont {Karimi},
  \citenamefont {Schulz}, \citenamefont {{De Leon}}, \citenamefont {Qassim},
  \citenamefont {Upham},\ and\ \citenamefont {Boyd}}]{Karimi2014}%
  \BibitemOpen
  \bibfield  {author} {\bibinfo {author} {\bibfnamefont {E.}~\bibnamefont
  {Karimi}}, \bibinfo {author} {\bibfnamefont {S.~A.}\ \bibnamefont {Schulz}},
  \bibinfo {author} {\bibfnamefont {I.}~\bibnamefont {{De Leon}}}, \bibinfo
  {author} {\bibfnamefont {H.}~\bibnamefont {Qassim}}, \bibinfo {author}
  {\bibfnamefont {J.}~\bibnamefont {Upham}}, \ and\ \bibinfo {author}
  {\bibfnamefont {R.~W.}\ \bibnamefont {Boyd}},\ }\href {\doibase
  10.1038/lsa.2014.48} {\bibfield  {journal} {\bibinfo  {journal} {Light:
  Science \& Applications}\ }\textbf {\bibinfo {volume} {3}},\ \bibinfo {pages}
  {e167} (\bibinfo {year} {2014})}\BibitemShut {NoStop}%
\bibitem [{\citenamefont {Hu}\ \emph {et~al.}(2013)\citenamefont {Hu},
  \citenamefont {Wang}, \citenamefont {Feng}, \citenamefont {Ye}, \citenamefont
  {Sun}, \citenamefont {Kan}, \citenamefont {Klar},\ and\ \citenamefont
  {Zhang}}]{Hu2013a}%
  \BibitemOpen
  \bibfield  {author} {\bibinfo {author} {\bibfnamefont {D.}~\bibnamefont
  {Hu}}, \bibinfo {author} {\bibfnamefont {X.}~\bibnamefont {Wang}}, \bibinfo
  {author} {\bibfnamefont {S.}~\bibnamefont {Feng}}, \bibinfo {author}
  {\bibfnamefont {J.}~\bibnamefont {Ye}}, \bibinfo {author} {\bibfnamefont
  {W.}~\bibnamefont {Sun}}, \bibinfo {author} {\bibfnamefont {Q.}~\bibnamefont
  {Kan}}, \bibinfo {author} {\bibfnamefont {P.~J.}\ \bibnamefont {Klar}}, \
  and\ \bibinfo {author} {\bibfnamefont {Y.}~\bibnamefont {Zhang}},\ }\href
  {\doibase 10.1002/adom.201200044} {\bibfield  {journal} {\bibinfo  {journal}
  {Advanced Optical Materials}\ }\textbf {\bibinfo {volume} {1}},\ \bibinfo
  {pages} {186} (\bibinfo {year} {2013})}\BibitemShut {NoStop}%
\bibitem [{\citenamefont {Pfeiffer}\ and\ \citenamefont
  {Grbic}(2013)}]{Pfeiffer2013}%
  \BibitemOpen
  \bibfield  {author} {\bibinfo {author} {\bibfnamefont {C.}~\bibnamefont
  {Pfeiffer}}\ and\ \bibinfo {author} {\bibfnamefont {A.}~\bibnamefont
  {Grbic}},\ }\href {\doibase 10.1103/PhysRevLett.110.197401} {\bibfield
  {journal} {\bibinfo  {journal} {Physical Review Letters}\ }\textbf {\bibinfo
  {volume} {110}},\ \bibinfo {pages} {197401} (\bibinfo {year}
  {2013})}\BibitemShut {NoStop}%
\bibitem [{\citenamefont {Monticone}\ \emph {et~al.}(2013)\citenamefont
  {Monticone}, \citenamefont {Estakhri},\ and\ \citenamefont
  {Al\`{u}}}]{Monticone2013}%
  \BibitemOpen
  \bibfield  {author} {\bibinfo {author} {\bibfnamefont {F.}~\bibnamefont
  {Monticone}}, \bibinfo {author} {\bibfnamefont {N.~M.}\ \bibnamefont
  {Estakhri}}, \ and\ \bibinfo {author} {\bibfnamefont {A.}~\bibnamefont
  {Al\`{u}}},\ }\href {\doibase 10.1103/PhysRevLett.110.203903} {\bibfield
  {journal} {\bibinfo  {journal} {Physical Review Letters}\ }\textbf {\bibinfo
  {volume} {110}},\ \bibinfo {pages} {203903} (\bibinfo {year}
  {2013})}\BibitemShut {NoStop}%
\bibitem [{\citenamefont {Balanis}(1989)}]{Balanis1989}%
  \BibitemOpen
  \bibfield  {author} {\bibinfo {author} {\bibfnamefont {C.~A.}\ \bibnamefont
  {Balanis}},\ }\href@noop {} {\emph {\bibinfo {title} {{Advanced Engineering
  Electromagnetics}}}}\ (\bibinfo  {publisher} {John Wiley \& Sons},\ \bibinfo
  {address} {New York},\ \bibinfo {year} {1989})\BibitemShut {NoStop}%
\bibitem [{\citenamefont {Collin}(1990)}]{Collin1990}%
  \BibitemOpen
  \bibfield  {author} {\bibinfo {author} {\bibfnamefont {R.~E.}\ \bibnamefont
  {Collin}},\ }\href
  {http://www.wiley.com/WileyCDA/WileyTitle/productCd-0879422378,miniSiteCd-IEEE2.html}
  {\emph {\bibinfo {title} {{Field Theory of Guided Waves}}}},\ \bibinfo
  {edition} {2nd}\ ed.\ (\bibinfo  {publisher} {Wiley-IEEE Press},\ \bibinfo
  {year} {1990})\BibitemShut {NoStop}%
\bibitem [{\citenamefont {McGrath}(1986)}]{McGrath1986}%
  \BibitemOpen
  \bibfield  {author} {\bibinfo {author} {\bibfnamefont {D.}~\bibnamefont
  {McGrath}},\ }\href {\doibase 10.1109/TAP.1986.1143726} {\bibfield  {journal}
  {\bibinfo  {journal} {IEEE Transactions on Antennas and Propagation}\
  }\textbf {\bibinfo {volume} {34}},\ \bibinfo {pages} {46} (\bibinfo {year}
  {1986})}\BibitemShut {NoStop}%
\bibitem [{\citenamefont {Stute}\ \emph {et~al.}(1996)\citenamefont {Stute},
  \citenamefont {Uniiwrsitj}, \citenamefont {Collejie},\ and\ \citenamefont
  {Pozar}}]{Pozar1996a}%
  \BibitemOpen
  \bibfield  {author} {\bibinfo {author} {\bibfnamefont {P.}~\bibnamefont
  {Stute}}, \bibinfo {author} {\bibfnamefont {S.}~\bibnamefont {Uniiwrsitj}},
  \bibinfo {author} {\bibfnamefont {S.}~\bibnamefont {Collejie}}, \ and\
  \bibinfo {author} {\bibfnamefont {D.}~\bibnamefont {Pozar}},\ }\href
  {\doibase 10.1049/el:19961451} {\bibfield  {journal} {\bibinfo  {journal}
  {Electronics Letters}\ }\textbf {\bibinfo {volume} {32}},\ \bibinfo {pages}
  {2109} (\bibinfo {year} {1996})}\BibitemShut {NoStop}%
\bibitem [{\citenamefont {Encinar}\ and\ \citenamefont
  {Zornoza}(2003)}]{Encinar2003}%
  \BibitemOpen
  \bibfield  {author} {\bibinfo {author} {\bibfnamefont {J.}~\bibnamefont
  {Encinar}}\ and\ \bibinfo {author} {\bibfnamefont {J.}~\bibnamefont
  {Zornoza}},\ }\href {\doibase 10.1109/TAP.2003.813611} {\bibfield  {journal}
  {\bibinfo  {journal} {IEEE Transactions on Antennas and Propagation}\
  }\textbf {\bibinfo {volume} {51}},\ \bibinfo {pages} {1662} (\bibinfo {year}
  {2003})}\BibitemShut {NoStop}%
\bibitem [{\citenamefont {Ryan}\ \emph {et~al.}(2010)\citenamefont {Ryan},
  \citenamefont {Chaharmir}, \citenamefont {Shaker}, \citenamefont {Bray},
  \citenamefont {Antar},\ and\ \citenamefont {Ittipiboon}}]{Ryan2010}%
  \BibitemOpen
  \bibfield  {author} {\bibinfo {author} {\bibfnamefont {C.~G.~M.}\
  \bibnamefont {Ryan}}, \bibinfo {author} {\bibfnamefont {M.~R.}\ \bibnamefont
  {Chaharmir}}, \bibinfo {author} {\bibfnamefont {J.}~\bibnamefont {Shaker}},
  \bibinfo {author} {\bibfnamefont {J.~R.}\ \bibnamefont {Bray}}, \bibinfo
  {author} {\bibfnamefont {Y.~M.~M.}\ \bibnamefont {Antar}}, \ and\ \bibinfo
  {author} {\bibfnamefont {A.}~\bibnamefont {Ittipiboon}},\ }\href {\doibase
  10.1109/TAP.2010.2044356} {\bibfield  {journal} {\bibinfo  {journal} {IEEE
  Transactions on Antennas and Propagation}\ }\textbf {\bibinfo {volume}
  {58}},\ \bibinfo {pages} {1486} (\bibinfo {year} {2010})}\BibitemShut
  {NoStop}%
\bibitem [{\citenamefont {Pozar}\ \emph {et~al.}(1997)\citenamefont {Pozar},
  \citenamefont {Targonski},\ and\ \citenamefont {Syrigos}}]{Pozar1997}%
  \BibitemOpen
  \bibfield  {author} {\bibinfo {author} {\bibfnamefont {D.}~\bibnamefont
  {Pozar}}, \bibinfo {author} {\bibfnamefont {S.}~\bibnamefont {Targonski}}, \
  and\ \bibinfo {author} {\bibfnamefont {H.}~\bibnamefont {Syrigos}},\ }\href
  {\doibase 10.1109/8.560348} {\bibfield  {journal} {\bibinfo  {journal} {IEEE
  Transactions on Antennas and Propagation}\ }\textbf {\bibinfo {volume}
  {45}},\ \bibinfo {pages} {287} (\bibinfo {year} {1997})}\BibitemShut
  {NoStop}%
\bibitem [{\citenamefont {Pozar}\ \emph {et~al.}(1999)\citenamefont {Pozar},
  \citenamefont {Targonski},\ and\ \citenamefont {Pokuls}}]{Pozar1999}%
  \BibitemOpen
  \bibfield  {author} {\bibinfo {author} {\bibfnamefont {D.}~\bibnamefont
  {Pozar}}, \bibinfo {author} {\bibfnamefont {S.}~\bibnamefont {Targonski}}, \
  and\ \bibinfo {author} {\bibfnamefont {R.}~\bibnamefont {Pokuls}},\ }\href
  {\doibase 10.1109/8.785748} {\bibfield  {journal} {\bibinfo  {journal} {IEEE
  Transactions on Antennas and Propagation}\ }\textbf {\bibinfo {volume}
  {47}},\ \bibinfo {pages} {1167} (\bibinfo {year} {1999})}\BibitemShut
  {NoStop}%
\bibitem [{\citenamefont {Pors}\ and\ \citenamefont
  {Bozhevolnyi}(2013)}]{Pors2013}%
  \BibitemOpen
  \bibfield  {author} {\bibinfo {author} {\bibfnamefont {A.}~\bibnamefont
  {Pors}}\ and\ \bibinfo {author} {\bibfnamefont {S.~I.}\ \bibnamefont
  {Bozhevolnyi}},\ }\href {\doibase 10.1364/OE.21.027438} {\bibfield  {journal}
  {\bibinfo  {journal} {Optics Express}\ }\textbf {\bibinfo {volume} {21}},\
  \bibinfo {pages} {27438} (\bibinfo {year} {2013})}\BibitemShut {NoStop}%
\bibitem [{\citenamefont {Verslegers}\ \emph {et~al.}(2009)\citenamefont
  {Verslegers}, \citenamefont {Catrysse}, \citenamefont {Yu}, \citenamefont
  {White}, \citenamefont {Barnard}, \citenamefont {Brongersma},\ and\
  \citenamefont {Fan}}]{Verslegers2009}%
  \BibitemOpen
  \bibfield  {author} {\bibinfo {author} {\bibfnamefont {L.}~\bibnamefont
  {Verslegers}}, \bibinfo {author} {\bibfnamefont {P.~B.}\ \bibnamefont
  {Catrysse}}, \bibinfo {author} {\bibfnamefont {Z.}~\bibnamefont {Yu}},
  \bibinfo {author} {\bibfnamefont {J.~S.}\ \bibnamefont {White}}, \bibinfo
  {author} {\bibfnamefont {E.~S.}\ \bibnamefont {Barnard}}, \bibinfo {author}
  {\bibfnamefont {M.~L.}\ \bibnamefont {Brongersma}}, \ and\ \bibinfo {author}
  {\bibfnamefont {S.}~\bibnamefont {Fan}},\ }\href {\doibase 10.1021/nl802830y}
  {\bibfield  {journal} {\bibinfo  {journal} {Nano letters}\ }\textbf {\bibinfo
  {volume} {9}},\ \bibinfo {pages} {235} (\bibinfo {year} {2009})}\BibitemShut
  {NoStop}%
\bibitem [{\citenamefont {Fattal}\ \emph {et~al.}(2010)\citenamefont {Fattal},
  \citenamefont {Li}, \citenamefont {Peng}, \citenamefont {Fiorentino},\ and\
  \citenamefont {Beausoleil}}]{Fattal2010}%
  \BibitemOpen
  \bibfield  {author} {\bibinfo {author} {\bibfnamefont {D.}~\bibnamefont
  {Fattal}}, \bibinfo {author} {\bibfnamefont {J.}~\bibnamefont {Li}}, \bibinfo
  {author} {\bibfnamefont {Z.}~\bibnamefont {Peng}}, \bibinfo {author}
  {\bibfnamefont {M.}~\bibnamefont {Fiorentino}}, \ and\ \bibinfo {author}
  {\bibfnamefont {R.~G.}\ \bibnamefont {Beausoleil}},\ }\href {\doibase
  10.1038/nphoton.2010.116} {\bibfield  {journal} {\bibinfo  {journal} {Nature
  Photonics}\ }\textbf {\bibinfo {volume} {4}},\ \bibinfo {pages} {466}
  (\bibinfo {year} {2010})}\BibitemShut {NoStop}%
\bibitem [{\citenamefont {Lu}\ \emph {et~al.}(2010)\citenamefont {Lu},
  \citenamefont {Sedgwick}, \citenamefont {Karagodsky}, \citenamefont {Chase},\
  and\ \citenamefont {Chang-Hasnain}}]{Lu2010}%
  \BibitemOpen
  \bibfield  {author} {\bibinfo {author} {\bibfnamefont {F.}~\bibnamefont
  {Lu}}, \bibinfo {author} {\bibfnamefont {F.~G.}\ \bibnamefont {Sedgwick}},
  \bibinfo {author} {\bibfnamefont {V.}~\bibnamefont {Karagodsky}}, \bibinfo
  {author} {\bibfnamefont {C.}~\bibnamefont {Chase}}, \ and\ \bibinfo {author}
  {\bibfnamefont {C.~J.}\ \bibnamefont {Chang-Hasnain}},\ }\href {\doibase
  10.1364/OE.18.012606} {\bibfield  {journal} {\bibinfo  {journal} {Optics
  express}\ }\textbf {\bibinfo {volume} {18}},\ \bibinfo {pages} {12606}
  (\bibinfo {year} {2010})}\BibitemShut {NoStop}%
\bibitem [{\citenamefont {Vo}\ \emph {et~al.}(2014)\citenamefont {Vo},
  \citenamefont {Fattal}, \citenamefont {Sorin}, \citenamefont {Peng},
  \citenamefont {Tran}, \citenamefont {Beausoleil},\ and\ \citenamefont
  {Fiorentino}}]{Vo2014}%
  \BibitemOpen
  \bibfield  {author} {\bibinfo {author} {\bibfnamefont {S.}~\bibnamefont
  {Vo}}, \bibinfo {author} {\bibfnamefont {D.}~\bibnamefont {Fattal}}, \bibinfo
  {author} {\bibfnamefont {W.~V.}\ \bibnamefont {Sorin}}, \bibinfo {author}
  {\bibfnamefont {Z.}~\bibnamefont {Peng}}, \bibinfo {author} {\bibfnamefont
  {T.}~\bibnamefont {Tran}}, \bibinfo {author} {\bibfnamefont {R.~G.}\
  \bibnamefont {Beausoleil}}, \ and\ \bibinfo {author} {\bibfnamefont
  {M.}~\bibnamefont {Fiorentino}},\ }\href {\doibase 10.1109/LPT.2014.2325947}
  {\bibfield  {journal} {\bibinfo  {journal} {IEEE Photonics Technology
  Letters}\ }\textbf {\bibinfo {volume} {26}},\ \bibinfo {pages} {1} (\bibinfo
  {year} {2014})}\BibitemShut {NoStop}%
\bibitem [{\citenamefont {Arbabi}\ \emph
  {et~al.}(2014{\natexlab{a}})\citenamefont {Arbabi}, \citenamefont {Bagheri},
  \citenamefont {Ball}, \citenamefont {Horie}, \citenamefont {Fattal},\ and\
  \citenamefont {Faraon}}]{Arbabi2014}%
  \BibitemOpen
  \bibfield  {author} {\bibinfo {author} {\bibfnamefont {A.}~\bibnamefont
  {Arbabi}}, \bibinfo {author} {\bibfnamefont {M.}~\bibnamefont {Bagheri}},
  \bibinfo {author} {\bibfnamefont {A.~J.}\ \bibnamefont {Ball}}, \bibinfo
  {author} {\bibfnamefont {Y.}~\bibnamefont {Horie}}, \bibinfo {author}
  {\bibfnamefont {D.}~\bibnamefont {Fattal}}, \ and\ \bibinfo {author}
  {\bibfnamefont {A.}~\bibnamefont {Faraon}},\ }in\ \href
  {http://www.opticsinfobase.org/abstract.cfm?URI=CLEO\_SI-2014-STu3M.4} {\emph
  {\bibinfo {booktitle} {CLEO: 2014}}}\ (\bibinfo  {publisher} {Optical Society
  of America},\ \bibinfo {address} {San Jose, California},\ \bibinfo {year}
  {2014})\ p.\ \bibinfo {pages} {STu3M.4}\BibitemShut {NoStop}%
\bibitem [{\citenamefont {Arbabi}\ \emph
  {et~al.}(2014{\natexlab{b}})\citenamefont {Arbabi}, \citenamefont {Horie},
  \citenamefont {Ball}, \citenamefont {Bagheri},\ and\ \citenamefont
  {Faraon}}]{Arbabi2014c}%
  \BibitemOpen
  \bibfield  {author} {\bibinfo {author} {\bibfnamefont {A.}~\bibnamefont
  {Arbabi}}, \bibinfo {author} {\bibfnamefont {Y.}~\bibnamefont {Horie}},
  \bibinfo {author} {\bibfnamefont {A.~J.}\ \bibnamefont {Ball}}, \bibinfo
  {author} {\bibfnamefont {M.}~\bibnamefont {Bagheri}}, \ and\ \bibinfo
  {author} {\bibfnamefont {A.}~\bibnamefont {Faraon}},\ }\href
  {http://arxiv.org/abs/1410.8261} {\  (\bibinfo {year}
  {2014}{\natexlab{b}})},\ \Eprint {http://arxiv.org/abs/1410.8261}
  {arXiv:1410.8261} \BibitemShut {NoStop}%
\bibitem [{\citenamefont {Arbabi}\ \emph
  {et~al.}(2014{\natexlab{b}})\citenamefont {Arbabi}, \citenamefont {Horie},
   \citenamefont {Bagheri},\ and\ \citenamefont
  {Faraon}}]{Arbabi2014d}%
  \BibitemOpen
  \bibfield  {author} {\bibinfo {author} {\bibfnamefont {A.}~\bibnamefont
  {Arbabi}}, \bibinfo {author} {\bibfnamefont {Y.}~\bibnamefont {Horie}},
  \bibinfo {author} {\bibfnamefont {M.}~\bibnamefont {Bagheri}}, \ and\ \bibinfo
  {author} {\bibfnamefont {A.}~\bibnamefont {Faraon}},\ }\href
  {http://arxiv.org/abs/1410.1411.1494} {\  (\bibinfo {year}
  {2014}{\natexlab{b}})},\ \Eprint {http://arxiv.org/abs/1411.1494}
  {arXiv:1411.1494} \BibitemShut {NoStop}%
\bibitem [{\citenamefont {Arbabi}\ \emph
  {et~al.}(2014{\natexlab{c}})\citenamefont {Arbabi}, \citenamefont {Horie},\
  and\ \citenamefont {Faraon}}]{Arbabi2014a}%
  \BibitemOpen
  \bibfield  {author} {\bibinfo {author} {\bibfnamefont {A.}~\bibnamefont
  {Arbabi}}, \bibinfo {author} {\bibfnamefont {Y.}~\bibnamefont {Horie}}, \
  and\ \bibinfo {author} {\bibfnamefont {A.}~\bibnamefont {Faraon}},\ }in\
  \href {http://www.opticsinfobase.org/abstract.cfm?URI=CLEO\_SI-2014-STu3M.5}
  {\emph {\bibinfo {booktitle} {CLEO: 2014}}}\ (\bibinfo  {publisher} {Optical
  Society of America},\ \bibinfo {address} {San Jose, California},\ \bibinfo
  {year} {2014})\ p.\ \bibinfo {pages} {STu3M.5}\BibitemShut {NoStop}%
\end{thebibliography}
\end{document}